\documentclass{ws-procs9x6}
\newcommand{\sq}{\lower.15ex\hbox{\large$\Box$}}

\begin{document}

\title{Gauge Invariance in 2PI Effective Actions}
\author{E. MOTTOLA}

\address{Theoretical Division, T-8 \\
MS B285, \\
Los Alamos National Laboratory\\
Los Alamos, NM 87545\\
E-mail: emil@lanl.gov}
\maketitle

\abstracts{The problem of maintaining gauge invariance in
the 2PI formulation of QED is discussed. A modified form of the
2PI effective action is suggested in which Ward identities for external
(background field) and internal (quantum field) gauge transformations
are both satisfied, but in different manners. The residual gauge-fixing
dependence in this modified 2PI formulation vanishes in a certain low momentum
limit, which allows it to be used reliably for calculating quantities such as
transport coeffcients and soft field relaxation in hot gauge theories.}

The 1PI (one particle irreducible) effective action is the standard form of the 
effective action, which is a functional of classical field variables whose 
values may be regarded as expectation values of the underlying quantum fields. 
The first variation of the 1PI effective action produces equations of motion for 
these expectation values. Although sufficient in principle for most purposes,
in practice the 1PI effective action is easily evaluated only in some approximation 
scheme, such as a loop expansion, Hartree or large $N$ expansion,\cite{CHKM} which 
limits its range of applicability. In particular, these simplest loop or mean field 
approximations are incapable of describing collisional relaxation processes in hot or 
dense non-equilibrium plasmas. A consistent treatment of the effects of the hard 
quasi-particle degrees of freedom in gauge theories on the more slowly varying mean 
fields, even to leading order in the gauge coupling, is an outstanding challenge for
non-equilibrium field theory. Yet a consistent formulation with gauge fields is just 
what is required in many applications, such as in formation and expansion of the 
quark-gluon plasma produced in an ultra-relativistic heavy ion collision, or in 
baryogenesis in a first order electroweak phase transition. Systematic improvements of 
the QCD equation of state beyond perturbation theory will also require such a 
consistent resummation scheme.

The 2PI (two particle irreducible) effective action is a general field theoretic
tool for addressing this problem.\cite{LW,CJT} In constructing it one introduces
sources for both the fields and their two-point bilinears. The Legendre transform
with respect to this extended set of sources results in an effective action
that is a functional of both mean fields and their two-point correlation functions,
treated on the same footing. The 2PI formulation, sometimes known also as the
`$\Phi$-derivable' effective action has the advantage of being thermodynamically
consistent, in the sense that any extremum of the 2PI free energy is automatically
a minimum if the underlying quantum Hamiltonian is bounded from below. An arbitrary
truncation scheme of $n$-point functions need not preserve this property in
general. Although the 2PI effective action becomes equal to the 1PI
effective action when evaluated at its exact extremum, somewhat different and
more efficient expansion methods suggest themselves in the 2PI formulation. For
example the proper self-energy, needed for evaluating collisional and damping
effects, and which connects the Boltzmann description with the field theoretic one may be
derived from the 2PI effective action from only the first few non-vanishing
two-loop skeleton diagrams.\cite{Ba} This corresponds to a quite non-trivial 
resummation of an infinite class of perturbative diagrams beyond the leading order 
large $N$ or Hartree approximations.

When the mean fields are gauge fields we face an additional problem.  The 2PI effective
action runs into trouble in its simplest form, since it concentrates on resumming 
two-point correlation functions but ignores vertex corrections, which must accompany 
those two-point functions in a gauge theory, in order for Ward identities to be 
satisfied. In this contribution I show that the simplest two-loop approximation to the 
2PI effective action in the simplest gauge theory, {\it i.e.} electromagnetism suffers
from this problem, but that the gauge non-invariance of a slightly modified
2PI formulation vanishes in a certain kinematic limit relevant for relaxational
processes.

If we apply the usual 2PI recipe to ordinary QED, treating the electron propagator
$\mathcal G$ and photon propagator ${\mathcal D}_{\mu\nu}$ on exactly the same footing
then we arrive at
\begin{eqnarray}
&&{\mathcal S}_{2PI} [A; {\mathcal G}, {\mathcal D}] = S_{cl}[A] -i {\rm Tr} \ln {\mathcal G}^{-1}
 + \frac{i}{2} {\rm Tr} \ln {\mathcal D}^{-1} \nonumber\\
&& \qquad -i {\rm Tr} \left(G^{-1}[A] \cdot {\mathcal G}
 - 1\right) + i {\rm Tr} \left(d^{-1} \cdot {\mathcal D} - 1\right) + \Phi_2[{\mathcal G
},
{\mathcal D}]\,.
\label{tPIact}
\end{eqnarray}
The classical action $S_{cl}[A]$ is the Maxwell action and the lowest order inverse
propagators for the electron and photon are defined by
\begin{equation}
G^{-1}[A] \equiv i\left(\gamma^{\mu}\partial_{\mu} - i\gamma^{\mu}A_{\mu} +m \right)
\,
\end{equation}
\begin{equation}
d^{-1}_{\mu\nu} \equiv - \frac{\delta^2 S_{cl}}{\delta A^{\mu} \delta A^{\nu}}
= - \frac{1}{e^2} \left(\eta_{\mu\nu} \sq -\partial_{\mu}\partial_{\nu}\right)\,.
\end{equation}
The functional (\ref{tPIact}) is perfectly gauge invariant under {\it background}
field gauge transformations,
\begin{equation}
A_{\mu} \rightarrow A_{\mu} + \partial_{\mu} \theta\,,
\label{Agtran}
\end{equation}
provided that the both the lowest order and exact electron inverse Green's functions
transform as
\begin{equation}
{\mathcal G}^{-1} \rightarrow e^{-i\theta}{\mathcal G}^{-1} e^{i\theta}
\label{Ggtran}
\end{equation}
and the functional $\Phi_2$ involves only traces. For example if $\Phi_2$ is
given by the lowest order non-trivial 2PI two-loop expression,
\begin{equation}
\Phi_2[{\mathcal G},{\mathcal D}] = \frac{1}{2} \int\,d^4x\ d^4x'\ {\mathcal D}_{\mu\nu}(x,x')\ 
{\rm tr} \left\{{\mathcal G}(x,x') \gamma^{\mu} {\mathcal G}(x', x) \gamma^{\nu}\right\}\,,
\label{phitwo}
\end{equation}
then invariance under ({\ref{Ggtran}) is assured by the cyclic property of the trace.

Despite this invariance under background field gauge transformations it is easy
to see that the 2PI effective action with $\Phi_2$ given by (\ref{phitwo}) cannot
be a fully satisfactory approximation to QED. The reason is that the photon self-energy
derived from $\Phi_2$, namely,
\begin{equation}
i\,{\rm tr} \left\{{\mathcal G}(x,x') \gamma_{\mu} {\mathcal G}(x', x) \gamma_{\nu}\right\}
\label{pol}
\end{equation}
is {\it not} transverse, in general, if ${\mathcal G} \neq G[A]$. Transversality
of the photon self-energy depends upon the Ward-Takahashi identity,
\begin{equation}
(p_1-p_2)_{\mu}\Gamma^{\mu}(p_1,p_2)  = {\mathcal G}^{-1} (p_1) - {\mathcal G}^{-1} (p_2)
\label{WTiden}
\end{equation}
between the photon-fermion vertex $\Gamma^{\mu}$ and the inverse fermion
propagators in the exact theory. Since we have a resummed Green's function
in the 2PI effective action but only the {\it bare} vertex function,
$\Gamma^{\mu}(p_1,p_2) = \gamma^{\mu}$, the identity (\ref{WTiden}) is not
satisfied, and the polarization (\ref{pol}) is not transverse.

The problem is that although the 2PI representation (\ref{tPIact}) is exact,
any truncation scheme for $\Phi_2$ in terms of skeleton diagrams involving
$\mathcal G$ and $\mathcal D$ leads to equations for these propagators that sum
only a certain subset of higher loop diagrams with a {\it fixed} topology,
determined by the choice of $\Phi_2$. However, the exact identity
(\ref{WTiden}) requires cancellations at a given loop order between diagrams
of {\it different} topology, not included in the same subset. This
non-invariance with respect to gauge transformations on {\it internal} lines
remains in any 2PI truncation scheme, nothwithstanding full invariance with
respect to {\it external} gauge fields $A_{\mu}$, which is guaranteed by the
general background field method. The difference between the Ward identities
satisfied by the expectation values and those typically {\it not} satisfied by the
internal propagators and vertices can be seen even in a non-gauge $O(N)$ scalar
$\phi^4$ theory, where the failure of Goldstone's theorem for the internal 
propagators was noted some time ago.\cite{BaGr} 

This problem may not be as severe as it appears at first sight.
After all, we are hoping to use any truncation scheme in the 2PI formalism
to calculate some quantities only to a given order in weak coupling
parameter. If we can show that the errors we make by not having a strictly
gauge invariant formulation are higher order in the coupling for the quantities
of interest, the truncation scheme based on the 2PI effective action would
still be useful and reliable to this order. Adopting such a strategy also
might make possible modifications of the original 2PI formalism by neglecting
some higher order contributions from the very beginning.

One such modification, suggested to restore the transversality of the photon
self-energy, is to replace the resummed polarization (\ref{pol}) by the lower
order form,
\begin{equation}
\Pi_{\mu\nu}(x,x') = i\,{\rm tr} \left\{G(x,x') \gamma_{\mu} G(x', x)
\gamma_{\nu}\right\}\,,
\label{polsim}
\end{equation}
so that the photon inverse propagator becomes
\begin{equation}
{\mathcal D}^{-1}_{\mu\nu} \rightarrow D^{-1}_{\mu\nu} \equiv d^{-1}_{\mu\nu} +
\Pi_{\mu\nu}
\,.
\label{photinv}
\end{equation}
Since this is nothing but the photon inverse propagator in the leading order 
of the large $N_f$ expansion, it is certainly transverse. Using the corresponding 
propagator $D_{\mu\nu}$ instead of ${\mathcal D}_{\mu\nu}$ everywhere in (\ref{tPIact}) 
and (\ref{phitwo}) results in the following form for the fermion inverse propagator 
and self-energy,
\begin{equation}
{\mathcal G}^{-1} = G^{-1} + \Sigma = G^{-1} -i \gamma^{\mu} {\mathcal G} \gamma^{\nu}
D_{\mu\nu}\,,
\label{ferm}
\end{equation}
which is still a non-trivial Schwinger-Dyson resummation for $\mathcal G$ beyond leading 
order. Varying this formula with respect to the background gauge field $A_{\mu}$, one 
obtains a non-trivial Schwinger-Dyson equation for the vertex,
\begin{equation}
\Gamma^{\mu} \equiv \frac{\delta {\mathcal G}^{-1}}{\delta A_{\mu}} = \gamma^{\mu} +
 i(\gamma^{\alpha} {\mathcal G} \Gamma^{\mu} {\mathcal G} \gamma^{\beta})\, D_{\alpha\beta}\,,
\end{equation}
which generates a sum over ladder diagrams.
Because this vertex function  is {\it defined} by varying ${\mathcal G}^{-1}$ with
respect to the background field, it  obeys the Ward identity (\ref{WTiden}). In this
way we may modify the original 2PI  effective action to satisfy the Ward identities for
both the internal vertex  $\gamma^{\mu}$, and the external resummed vertex
$\Gamma^{\mu}$, albeit in different  ways.

It might seem that we have avoided the gauge non-invariance problem entirely by
this method. This is not the case, however, because gauge invariance also
requires that in inverting the transverse $D^{-1}_{\mu\nu}$ of (\ref{photinv}) to find 
the photon propagator $D_{\mu\nu}$ to appear in (\ref{ferm}), we should be allowed to add
an arbitrary longitudinal part to $D_{\mu\nu}$ without affecting the results. Because
the self-energy $\Sigma$ in (\ref{ferm}) involves a resummed fermion propagator
$\mathcal G$ but only the bare vertex $\gamma^{\mu}$, one can see fairly easily that the
usual proof that these longitudinal gauge dependent pieces of $D_{\mu\nu}$ drop
out will {\it not} go through. In other words, the fermion self-energy is still
gauge-fixing dependent.\cite{AS} It is again higher loop diagrams with a different
topology (crossed rainbows) that are required, but which are not included in the
resummation of rainbow diagrams given by (\ref{ferm}). 

If one does go through the usual proof of gauge fixing independence, one finds that the
non-cancellation within the rainbow approximation generated by (\ref{ferm}) 
{\it vanishes} in the limit that the fermion momentum $p$ is much larger than the
photon momentum $k$. Indeed the non-cancellation is of order 
$\vert \frac{k^2}{p^2}\vert$. Hence observables sensitive only to the
kinematic range of momenta $\vert k^2\vert \ll \vert p^2\vert$ may be reliably 
computed by the modified 2PI scheme,and are gauge invariant up to corrections of this 
order. This kinematic regime of soft photon exchange is precisely that which determines 
the zero frequency electrical conductivity of the high temperature QED plasma.\cite{Je} 
It is also the kinematic region of interest in the gap equation for the fermions
in high density QCD.\,\cite{PR} Further, by including higher order skeleton diagrams in
$\Phi_2$ in this modified 2PI formulation one may systematically suppress the
residual gauge dependence to higher orders in $\vert \frac{k^2}{p^2}\vert$. A
proof of these bounds and application to relaxational problems in non-equilibrium
gauge theories will be presented elsewhere. 

This research is supported by the Department of Energy, under contract
{\it W-7405-ENG-36} and is LANL Technical Report, {\it LA-UR-03-2804}.

\end{document}